\documentclass[%
 reprint,
nofootinbib,
 amsmath,amssymb,
 aps,
]{revtex4-2}

\usepackage{graphicx}
\usepackage{dcolumn}
\usepackage{amssymb}
\usepackage{lipsum}
\usepackage{amsmath}
\usepackage{bm} 

\begin{document}


\title[]{Unregulated Divergences of Feynman Integrals}

\author{Wen Chen}
 \email{chenwenphy@gmail.com}
\affiliation{ Key Laboratory of Atomic and Subatomic Structure and Quantum Control (MOE), Guangdong Basic Research Center of Excellence for Structure and Fundamental Interactions of Matter, Institute of Quantum Matter, South China Normal University, Guangzhou 510006, China
}
\affiliation{Guangdong-Hong Kong Joint Laboratory of Quantum Matter, Guangdong Provincial Key Laboratory of Nuclear Science, Southern Nuclear Science Computing Center, South China Normal University, Guangzhou 510006, China}
\date{\today}

\begin{abstract}
Feynman integrals can be expanded asymptotically with respect to some small parameters at the integrand level, a technique known as the expansion by regions. A naive expansion by regions may break down due to divergences not regulated by the spacetime dimension, exemplified by the rapidity divergences. A criterion to identify unregulated divergences is provided in this article. The analysis is conducted using both parametric and Mellin-Barnes representations, leading to a consistent conclusion. Based on this analysis, it is proven that {\it the presence of unregulated divergences implies the degeneracies of regions}.
\end{abstract}


\maketitle




\section{Introduction}
The investigation of the asymptotic expansions of Feynman amplitudes has a long history~\cite{Weinberg:1959nj,Zimmermann:1972tv,Chetyrkin:1988zz,Chetyrkin:1988cu,Smirnov:1990rz,Tkachov:1991ev,Pivovarov:1991du}. The modern approach to the asymptotic expansions is the method of regions~\cite{Beneke:1997zp,Smirnov:1999bza,Smirnov:2002pj,Pak:2010pt,Jantzen:2011nz}, according to which, a Feynman integral can be expanded as a series of a small parameter at the integrand level according to a prescribed scaling in each region. The sum of contributions of all regions reproduces the full integral. By factorizing different scales, the method of regions highly simplifies the calculations of Feynman integrals. Furthermore, it is intimately related to some effective theories~\cite{Beneke:1997zp}, such as heavy quark effective theory~(HQET)~\cite{Isgur:1989vq,Isgur:1990yhj}, nonrelativistic quantum chromodynamics~(NRQCD)~\cite{Caswell:1985ui,Bodwin:1994jh}, and soft collinear effective theory~(SCET)~\cite{Bauer:2000yr,Bauer:2001yt}. There are correspondences between the contributions of regions and the modes in effective theories.

Each term of the asymptotic expansion of an integral is homogeneous in the small parameter. Thus, it is expected that the method of regions may break down if an integral has an explicit logarithmic dependence. In some circumstances, a naive asymptotic expansion indeed results in some ill-defined integrals due to the presence of divergences not regulated by the spacetime dimension, as was first noticed in ref.~\cite{Collins:1992tv}~\footnote{I am grateful to M.~Kalmykov for bringing this paper and some relevant references to my attention.}. Examples are the pinch singularities in HQET and the rapidity divergences in SCET. Unregulated divergences lead to various problems. For example, on the theoretical side, a naive collinear factorization breaks down due to these divergences~\cite{Beneke:2003pa,Manohar:2006nz,Chiu:2007yn,Becher:2010tm}. On the technical side, auxiliary regulators need to be introduced~\cite{Collins:1989bt,Collins:1992tv,Usyukina:1992jd,Smirnov:1997gx,Smirnov:1998vk,Becher:2011dz,Li:2016axz}, which may highly complicate the calculations. In recent years, unregulated divergences, particularly rapidity divergences, have garnered significant interest due to their frequent occurrence in the calculation of power corrections~\cite{Ebert:2018gsn,Moult:2019uhz,Moult:2019vou,Liu:2019oav,Liu:2020tzd,Inglis-Whalen:2021bea,Vladimirov:2021hdn,Bell:2022ott,delCastillo:2023rng,Mukherjee:2023snp,Ferrera:2023vsw,Beneke:2024cpq}. Consequently, a deeper understanding of unregulated divergences is highly desirable.

In a recent paper~\cite{Chen:2023hmk}, a method for the recursive calculations of Feynman integrals was developed. Specifically, a delta function $\delta(y-\frac{x_i}{x_j})$, with $x_i$ Feynman parameters, is inserted into a parametric integral. The obtained integral is calculated by using the differential equation method~\cite{Kotikov:1990kg,Remiddi:1997ny}. The boundary conditions of the differential equations are expressed in terms of parametric integrals which can further be calculated by using this method. However, an arbitrary choice of the delta function may result in boundary integrals with unregulated divergences. Thus the recursive procedure breaks down. Hence, it is necessary to have a criterion to identify unregulated divergences while choosing the delta function.

A criterion based on the Lee-Pomeransky representation~\cite{Lee:2013hzt} was provided in ref.~\cite{Heinrich:2021dbf}~\footnote{I am grateful to G.~Heinrich for bringing this reference to my attention.}. In this article, an equivalent criterion is given. The consequence of unregulated divergences on the asymptotic expansions is studied.

This paper is organized as follows. In sec.~\ref{sec:Exampl}, we use a simple vertex integral to show how the unregulated divergences arise. In sec.~\ref{sec:UnrDiv}, based on the parametric representation, we provide a criterion to identify unregulated divergences. We also analyze the divergences in the Mellin-Barnes representation and get a consistent conclusion. In sec.~\ref{sec:DegReg}, we further prove that the presence of unregulated divergences always implies the degeneracies of regions. Based on these general methods, we reanalyze the vertex integral example in sec.~\ref{sec:ReanExampl}.

\section{A pedagogical example}\label{sec:Exampl}
We start with a simple vertex integral
\begin{equation}\label{eq:Exampl}
    J_0(i_1,~i_2,~i_3)\equiv\int\frac{\mathrm{d}^dl}{\pi^{d/2}}\frac{1}{(l+p_1)^{2i_1}(l-p_2)^{2i_2}\left(l^2-m^2\right)^{i_3}}~,
\end{equation}
with $p_i^2=0$ and $p_1\cdot p_2=\frac{1}{2}Q^2$. We use the light cone coordinates and denote $l^+\equiv l\cdot p_1$ and $l^-\equiv l\cdot p_2$.

We are interested in the small mass limit $m\ll Q$. Thus we consider the asymptotic expansion of this integral with respect to $\frac{m}{Q}$. For simplicity, we take $Q^2=2$ hereafter. There are three regions:
\begin{align}\label{eq:Reg}
    \text{collinear a:}&\qquad l^+\sim m^0,~l^-\sim l^2\sim m^2~,\nonumber\\
    \text{collinear b:}&\qquad l^-\sim m^0,~l^+\sim l^2\sim m^2~,\\
    \text{hard:}&\qquad l^+\sim l^-\sim l^2\sim m^0~.\nonumber
\end{align}
We focus on $J_0(1,~1,~1)$ and consider the contribution of the region ``collinear a'', in which the integral $J_0$ is reduced to
\begin{equation}\label{eq:IntJa}
    J_a=\int\frac{\mathrm{d}^dl}{\pi^{d/2}}\frac{1}{2l^+(l-p_2)^2\left(l^2-m^2\right)}~.
\end{equation}
According to the power counting of the region ``collinear a'', it is easy to get the scaling $J_a\sim m^{-2\epsilon}$. This integral is divergent. The divergence arises from the singularity at $l^+=0$, which is not regulated by the space-time dimension $d$, since the denominator $2l^+$ is free of the transverse momentum. We can also analyze the divergence through the Feynman parameter representation
\begin{equation}
\begin{split}
    J_a=&-i~\Gamma(1+\epsilon)\int_0^\infty\mathrm{d}x_1\mathrm{d}x_2\mathrm{d}x_3~\delta(1-x_2-x_3)\\
    &\times(x_2+x_3)^{-1+2\epsilon}\left[m^2x_3(x_2+x_3)-2x_1x_2\right]^{-1-\epsilon}\\
    =&-i~\Gamma(1+\epsilon)\int_0^\infty\mathrm{d}x_1\int_0^1\mathrm{d}x_2\left[m^2(1-x_2)-2x_1x_2\right]^{-1-\epsilon}~,
\end{split}
\end{equation}
where $\epsilon=\frac{1}{2}(4-d)$, and the Feynman parameters $x_i$ are in correspondence with the propagators in eq.~(\ref{eq:IntJa}). The unregulated divergence arises from the region where $x_2\to 0$ and $x_1\sim x_2^{-1}$. Rescaling $x_1$ by $x_1\to\frac{x_1}{x_2}$, we get
\begin{equation}
\begin{split}
    J_a=&-i\Gamma(1+\epsilon)\int_0^\infty\mathrm{d}x_1\int_0^1\mathrm{d}x_2\frac{1}{x_2}\left[m^2(1-x_2)-2x_1\right]^{-1-\epsilon}~.
\end{split}
\end{equation}
We see that the singularity at $x_2=0$ is not regulated by $\epsilon$.

Since the original integral $J_0$ is well-defined, the divergence must arise from the asymptotic expansion. To see what is going on, instead of expanding the propagators in power series, we do a Mellin transformation and get
\begin{equation}
\begin{split}
    J_0(1,~1,~1)=&\int_{-i\infty}^{i\infty}\frac{\mathrm{d}z}{2\pi i}\Gamma(-z)\Gamma(1+z)\\
    &\times\int\frac{\mathrm{d}^dl}{\pi^{d/2}}\frac{(l^2)^z}{\left(l^2-m^2\right)(2l^+)^{1+z}(l-p_2)^2}\\
    \equiv&\int_{-i\infty}^{i\infty}\frac{\mathrm{d}z}{2\pi i}\Gamma(-z)\Gamma(1+z)J_{a}(z)~.
\end{split}
\end{equation}
The asymptotic expansion in the region ``collinear a'' is equivalent to picking the residue of the integrand at $z\in\mathbb{N}$ assuming that $J_a(z)$ is regular. However, as we have already known, $J_a(z)$ is singular at least at $z=0$. Thus a naive asymptotic expansion gives a divergent integral. Taking the singularity of $J_a(z)$ at $z=0$ into consideration, we get the contribution of the collinear regions
\begin{equation}
\begin{split}
    J_{ab}=&-\mathrm{Res}_{z=0}\left\{\Gamma(-z)\Gamma(1+z)J_{a}(z)\right\}+\mathcal{O}(m)\\
    =&\frac{i\Gamma(\epsilon)}{2}m^{-2\epsilon}\left[-2 \log (m)+\psi(\epsilon )-2 \psi(1-\epsilon ) -\gamma\right.\\
    &\left.+\log (2)-i \pi\right]+\mathcal{O}(m)~.
\end{split}
\end{equation}
Here $\psi(x)$ is the digamma function, and $\gamma_E$ is the Euler constant. The subscript $ab$ indicates that $J_{ab}$ is in fact the sum of contributions of the regions ``collinear a'' and ``collinear b''. This can easily be checked by introducing an auxiliary regulator.

We see that $J_{ab}$ has an explicit $\log(m)$ dependence in $d$ dimensions. We can further confirm this by using the differential-equation method. We choose the basis
\begin{equation}
    J=\{J_0(0,2,1),~J_0(2,0,0),~\epsilon J_0(1,1,1)\}~.
\end{equation}
Approximately, we have
\begin{equation}
\frac{\mathrm{d}J}{\mathrm{d}m}\approx \frac{1}{m}M\cdot J~,
\end{equation}
with
\begin{equation}\label{eq:ResMatr}
M=\epsilon~\begin{pmatrix}
 0 & 0 & 0 \\
 0 & -2 & 0 \\
 -2 & -1 & -2 \\
\end{pmatrix}~.
\end{equation}
$M$ has two eigenvalues: $\{0,~-2\epsilon\}$. The latter has a degree of degeneracy $2$. As a consequence, the asymptotic solutions of the differential equations take the form (see e.g. ref.~\cite{Ince:1956})
\begin{equation}\label{eq:AsymptSol}
J_0(1,1,1)\approx c_1+c_2m^{-2\epsilon}+c_3\log(m)m^{-2\epsilon}~.
\end{equation}
Again, we get an explicit $\log(m)$ dependence.

\section{Unregulated divergences}\label{sec:UnrDiv}
\subsection{Parametric representation}\label{sec:ParRepr}
We consider a general parametric integral of the representation~\cite{Chen:2019mqc,Chen:2019fzm}
\begin{equation}\label{eq:ParInt}
\begin{split}
I(\lambda_0,\lambda_1,\ldots,\lambda_n)=&\frac{\Gamma(-\lambda_0)}{\prod_{i=1}^{n+1}\Gamma(\lambda_i+1)}\int \mathrm{d}\Pi^{(n+1)}\mathcal{F}^{\lambda_0}\prod_{i=1}^{n+1}x_i^{\lambda_i}~.
\end{split}
\end{equation}
Here the integration measure is $\mathrm{d}\Pi^{(n+1)}\equiv\prod_{i=1}^{n+1}\mathrm{d}x_i\delta(1-\mathcal{E}^{(1)}(x))$, with $\mathcal{E}^{(1)}(x)$ a positive definite homogeneous function of $x$ of degree $1$. $\mathcal{F}$ is a homogeneous polynomial of $x$. For a $L$-loop integral with $n$ propagators, $\mathcal{F}$ is of degree $L+1$. It is related to the well-known Symanzik polynomials $U$ and $F$ through $\mathcal{F}=F+Ux_{n+1}$. $\lambda_0$ is related to the spacetime dimension $d$ through $d=-2\lambda_0$.

A general $\mathcal{F}$ polynomial is of the structure
\begin{equation}\label{FPol}
\mathcal{F}=\sum_{a=1}^{A}\left(C_{\mathcal{F},a}\prod_{i=1}^{n+1}x_i^{\Lambda_{ai}}\right),
\end{equation}
The idea of regions and the geometric method described in ref.~\cite{Pak:2010pt} can be generalized to general parametric integrals without referring to a physical scale (see e.g. ref.~\cite{Heinrich:2021dbf}). That is, by a region, we mean a subset $S_r$ of $\{1,~2,\dots,~A\}$ together with a vector $\bm{k}_r$ (called a region vector) such that\footnote{Generally speaking, we have $\sum_{i=1}^{n+1}\Lambda_{ai}k_{r,i}\geq b_r$ for some constant $b_r$, but we can always make it vanish by virtue of the homogeneity of $\mathcal{F}$.}
\begin{subequations}\label{eq:RegDef}
    \begin{align}
        \sum_{i=1}^{n+1}\Lambda_{ai}k_{r,i}=&0,\quad a\in S_r~,\\
        \sum_{i=1}^{n+1}\Lambda_{ai}k_{r,i}>&0,\quad a\notin S_r~.
    \end{align}
\end{subequations}
We normalize $\bm{k}_r$ such that
\begin{equation}
    \min_{a\notin S_r}\left\{\sum_ik_{r,i}\Lambda_{ai}\right\}=1~.
\end{equation}
Since all the elements of $\Lambda$ are integers, all the components of $\bm{k}_r$ are rational numbers. Obviously, terms in $S_r$ dominate $\mathcal{F}$ when the Feynman parameters scale as $x_i\sim s^{k_{r,i}}$ with respect to an artificial scale $s$. The delta function in the measure $\mathrm{d}\Pi^{(n+1)}$ is of order $1$. According to eq.~(\ref{eq:RegDef}), the scaling of the $\mathcal{F}$ polynomial is $\mathcal{F}\sim s^{\sum_{i=1}^{n+1}\Lambda_{ai}k_{r,i}}=s^0$. Thus it is easy to see that in the region $r$ the integral $I(\lambda_0,\lambda_1,\ldots,\lambda_n)$ scales as
\begin{equation}
    I(\lambda_0,\lambda_1,\ldots,\lambda_n)\sim s^{\sum_{i=1}^{n+1}k_{r,i}(\lambda_i+1)}\equiv s^{\nu_r}~.
\end{equation}
Notice that $\nu_r$ depends on the spacetime dimension $d$ through the $d$-dependence of $\lambda_{n+1}$.

While the scale $s$ is artificial, the scaling $\nu_r$ is meaningful. To see this, we insert a trivial integral $\int\mathrm{d}y~\delta(y-\frac{x_i}{x_j})$, with $k_{r,i}>k_{r,j}$, into the parametric integral $I$, as in ref.~\cite{Chen:2023hmk}. Then we convert the parametric integral to an integral with respect to $y$. That is, $I=\int_0^\infty\mathrm{d}y~I_y$, where $I_y$ is a $y$-dependent integral. It can be shown that there are correspondences between the region vectors $\bm{k}_r$ and the regions of the integral $I_y$ (with respect to the asymptotic expansion in $y\to 0$). And in the region corresponding to $\bm{k}_r$, we have $I_y\sim y^{\frac{\nu_r}{k_{r,i}-k_{r,j}}-1}$.~(For the details, see sec.~3.2 of ref.~\cite{Chen:2023hmk}.) Thus the integral $\int_0^\infty\mathrm{d}y~I_y$ is divergent if $\nu_r\leq0$. This divergence is not regulated by the spacetime dimension if $\nu_r$ is an integer~\footnote{If $\nu_r$ is a negative fractional number, we may introduce a regulator first and safely remove the regulator after the integration.}. Thus, we have the following criterion:

{\it A parametric integral $I(\lambda_0,\lambda_1,\ldots,\lambda_n)$ defined in eq.~(\ref{eq:ParInt}) is singular if there is a region $r$ such that
\begin{equation}\label{eq:Crit}
    \nu_r\equiv\sum_{i=1}^{n+1}k_{r,i}(\lambda_i+1)\in\mathbb{Z}^-\cup\{0\}~,
\end{equation}
with $k_r$ defined by eq.~(\ref{eq:RegDef}).}

\subsection{Mellin-Barnes representation}
By doing a $A-(n+1)$ fold Mellin transformation, we get
\begin{align*}
&I(\lambda_0,\lambda_1,\ldots,\lambda_n)\\
=&\frac{1}{\prod_{i=1}^{n+1}\Gamma(\lambda_i+1)}\int\prod_{a=n+2}^A\frac{\mathrm{d}z_a}{2\pi i}~\Gamma\left(-\lambda_0^\prime\right)\prod_{a=n+2}^{A}\Gamma(-z_a)C_{\mathcal{F},a}^{z_a}\\
&\times\int \mathrm{d}\Pi^{(n+1)}~\left[\sum_{a=1}^{n+1}\left(C_{\mathcal{F},a}\prod_{i=1}^{n+1}x_i^{\Lambda_{ai}}\right)\right]^{\lambda_0^\prime}\prod_{i=1}^{n+1}x_i^{\lambda_i^\prime}~,
\end{align*}
where $\lambda_0^\prime\equiv\lambda_0-\sum_{a=n+2}^{A}z_a$, $\lambda_i^\prime\equiv\lambda_i+\sum_{a=n+2}^Az_a\Lambda_{ai}$, and $C_{\mathcal{F},a}$ are those defined in eq.~(\ref{FPol}). As was shown in ref.~\cite{Chen:2023hmk}, an integral of which the $\mathcal{F}$ polynomial has exactly $n+1$ terms can be expressed in terms of gamma functions. That is,
\begin{equation}
\begin{split}
&\int\mathrm{d}\Pi^{(n+1)}\left[\sum_{a=1}^{n+1}\left(C_{\mathcal{F},a}\prod_{i=1}^{n+1}x_i^{\Lambda_{ai}}\right)\right]^{\lambda_0^\prime}\prod_{i=1}^{n+1}x_i^{\lambda_i^\prime}\\
=&\frac{(L+1)\prod_{a=1}^{n+1}\left[\Gamma(\bar{\lambda}_a^\prime)C_{\mathcal{F},a}^{-\bar{\lambda}_a^\prime}\right]}{\parallel\bar{\Lambda}\parallel\Gamma(-\lambda_0^\prime)},
\end{split}
\end{equation}
where $\bar{\lambda}_a^\prime=\sum_{i=1}^{n+1}(\Lambda^{\prime-1})_{ia}(\lambda_i^\prime+1)$, with $\bar{\Lambda}$ a squared matrix such that $\bar{\Lambda}_{ai}=\Lambda_{ai},~a\leq n+1$. Then, we get
\begin{equation}\label{eq:MBInt}
\begin{split}
&I(\lambda_0,\lambda_1,\ldots,\lambda_n)\\
=&\frac{(L+1)}{\parallel\bar{\Lambda}\parallel\prod_{i=1}^{n+1}\Gamma(\lambda_i+1)}\int\prod_{a=n+2}^A\frac{\mathrm{d}z_a}{2\pi i}\\
&\prod_{a=1}^{n+1}\left[\Gamma(\bar{\lambda}_a^\prime)C_{\mathcal{F},a}^{-\bar{\lambda}_a^\prime}\right]\prod_{a=n+2}^{A}\left[\Gamma(-z_a)C_{\mathcal{F},a}^{z_a}\right]\\
=&\frac{L+1}{\prod_{i=1}^{n+1}\Gamma(\lambda_i+1)}\int\prod_{a=1}^A\frac{\mathrm{d}z_a}{2\pi i}\prod_{j=1}^{n+1}\left[2\pi i\delta(Z_j)\right]\prod_{a=1}^A
\left[C_{\mathcal{F},a}^{z_a}\Gamma(-z_a)\right],
\end{split}
\end{equation}
where $Z_i\equiv\lambda_{i}+1+\sum_{a=1}^{A}z_a\Lambda_{ai}$.

It was shown in ref.~\cite{Ananthanarayan:2020fhl} that a Mellin-Barnes integral has a multiple-series representation. Here we do not need the full method. We only need to know that the integral in eq.~(\ref{eq:MBInt}) can be expressed in terms of a sum of building blocks of the form
\begin{equation}\label{eq:BuildBlock}
\begin{split}
    B_C=&\frac{L+1}{\prod_{i=1}^{n+1}\Gamma(\lambda_i+1)}\\
    &\times\sum_{n_a=0}^\infty\left\{\prod_{a\in C}
\left[C_{\mathcal{F},a}^{\bar{z}_a}\Gamma(-\bar{z}_a)\right]\prod_{a\notin C}
\frac{(-1)^{n_a}C_{\mathcal{F},a}^{n_a}}{n_a!}\right\}~,
\end{split}
\end{equation}
where $C$ is a subset of $\{1,2,\dots,~A\}$ of length $n+1$, and $\bar{z}_a\equiv-\sum_{i\in C}V_{ia}\left(\lambda_i+1+\sum_{b\notin C}n_b\Lambda_{bi}\right)$, with $V$ a matrix such that $\sum_{a\in C}\Lambda_{ia}V_{aj}=\delta_{ij}$.

Suppose that there is a region $r$ such that the criterion of eq.~(\ref{eq:Crit}) is satisfied. Without loss of generality, we assume that $\sum_ik_{r,i}\Lambda_{1i}\leq\sum_ik_{r,i}\Lambda_{ai},~1,\,a\notin S_r$. According to the normalization of $\bm{k}_r$, we have $\sum_ik_{r,i}\Lambda_{1i}=1$. We choose a series representation such that there is a block $C$ such that $1$ is an element of $C$, and all the rest elements of $C$ are in $S_r$. This representation is consistent with the asymptotic expansion in the region $r$. Obviously we have $V_{1i}=k_{r,i}$. Hence, $\bar{z}_1=-\nu_r-\sum_{i=1}^{n+1}\sum_{a\notin C}k_{r,i}n_a\Lambda_{ai}$. For the leading term of the series $B_C$ in eq.~(\ref{eq:BuildBlock}), we have $n_a=0$. Then $\bar{z}_1=-\nu_r\in\mathbb{Z}^+\cup\{0\}$, which makes $\Gamma(-\bar{z}_1)$ singular. Thus the leading term of $B_C$ is divergent, and so is the whole series~\footnote{A flaw of this argument is that it is not certain whether this divergence is essential to the integral or specific to the series representation we choose.}. In this case, some of the singularities of the Mellin-Barnes integral are pinched and it falls into the resonant case~\cite{Ananthanarayan:2020fhl}. In the resonant case, the series representation has a logarithmic dependence. As was mentioned before, a logarithmic dependence can not be reproduced by the method of regions (without introducing auxiliary regulators).

\section{Degeneracies of regions}\label{sec:DegReg}
Correspondences can be built between the regions and the asymptotic solutions of differential equations of Feynman integrals. An observation from the asymptotic solutions of differential equations is that the asymptotic solutions may have an explicit logarithmic dependence if some regions are degenerate. Here, by degenerate, we mean that the leading powers of two regions are the same or differ by a rational number (rather than an irrational number or a function of the regulators). As can be seen from the example shown in sec.~\ref{sec:Exampl}, the two collinear regions share the same leading power $-2\epsilon$, and thus are degenerate. Correspondingly, the matrix $M$ in eq.~(\ref{eq:ResMatr}) has an eigenvalue $-2\epsilon$ with a multiplicity $2$. Consequently, the coefficient of $m^{-2\epsilon}$ of the asymptotic solution in eq.~(\ref{eq:AsymptSol}) has an explicit $\log m$ dependence. This observation indicates that the presence of unregulated divergences is always accompanied by the degeneracies of regions. In this section, we will prove that this is indeed true.

We consider the asymptotic expansion of the parametric integral $I(\lambda_0,\lambda_1,\ldots,\lambda_n)$ with respect to a kinematic variable, denoted by $x_0$. We assume that this integral is well-defined (before the asymptotic expansion). That is, it is free of unregulated divergences. For this integral, we define
\begin{equation}
\mathcal{F}=\sum_{a=1}^{A}\left(C_{\mathcal{F},a}^\prime\prod_{i=0}^{n+1}x_i^{\Lambda_{ai}^\prime}\right)~.
\end{equation}
Obviously we have $\Lambda_{ai}^\prime=\Lambda_{ai},~i\neq0$, and $C_{\mathcal{F},a}=x_0^{\Lambda_{a0}}C_{\mathcal{F},a}^\prime$. For a region $\rho$, we have a set $S_\rho$ and a vector $\bm{k}_\rho$ with $k_{\rho,0}=1$ such that
\begin{subequations}
    \begin{align}
        \sum_{i=0}^{n+1}\Lambda_{ai}^\prime k_{\rho,i}=&b_\rho,\quad a\in S_\rho~,\\
        \sum_{i=0}^{n+1}\Lambda_{ai}^\prime k_{\rho,i}>&b_\rho,\quad a\notin S_\rho~.
    \end{align}
\end{subequations}
To distinguish $\bm{k}_\rho$ from the region vectors $\bm{k}_r$ defined in sec.~\ref{sec:ParRepr}, we call $\bm{k}_\rho$ a scaling vector in this paper. For future convenience, we assume that $\Lambda_{a0}^\prime=0,~a\in S_\rho$. We can always make this true by rescaling all the Feynman parameters and factoring out a global factor out of $\mathcal{F}$ by virtue of the homogeneity of $\mathcal{F}$. It is easy to get the scaling of the integral $I(\lambda_0,\lambda_1,\ldots,\lambda_n)$ in the region $\rho$.
\begin{equation}
    I(\lambda_0,\lambda_1,\ldots,\lambda_n)\sim x_0^{\nu_\rho}~,
\end{equation}
where $\nu_\rho=b_\rho\lambda_0+\sum_{i=1}^{n+1}k_{\rho,i}(\lambda_{i}+1)$.

Suppose that the integral obtained after the asymptotic expansion in a region $\rho$ is singular. Then, there is a set $S_{\rho,r}\subset S_\rho$ and a vector $\bm{k}_{\rho,r}$, such that
\begin{subequations}
    \begin{align}
        &\sum_{i=1}^{n+1}\Lambda_{ai}k_{\rho,r,i}=0,\quad a\in S_{\rho,r}\subset S_\rho~,\\
        &\sum_{i=1}^{n+1}\Lambda_{ai}k_{\rho,r,i}>0,\quad a\in S_\rho~,a\notin S_{\rho,r},~\\
        &\nu_{\rho,r}\equiv\sum_{i=1}^{n+1}k_{\rho,r,i}(\lambda_i+1)\in\mathbb{Z}^-\cup\{0\}~.
    \end{align}
\end{subequations}
Since the original integral $I(\lambda_0,\lambda_1,\ldots,\lambda_n)$ is well defined, $S_{\rho,r}$ should not be a region of it. Thus, there is a subset $T_{\rho,r}$ of the complement of $S_\rho$(denoted by $\bar{S}_\rho$), such that
\begin{equation}
    \sum_{i=1}^{n+1}\Lambda_{ai}k_{\rho,r,i}<0,\quad a\in T_{\rho,r}\subset \bar{S}_\rho~.
\end{equation}

Now we introduce a vector $\bm{k}_{\rho^\prime}$, such that $k_{\rho^\prime,0}=k_{\rho,0}$, and $k_{\rho^\prime,i}=k_{\rho,i}+ck_{\rho,r,i},~i\neq0$, with $c$ a constant. Because, for an $a\in T_{\rho,r}$, $\sum_{i=0}^{n+1}\Lambda_{ai}^\prime k_{\rho^\prime,i}>b_\rho$ when $c=0$, and $\sum_{i=0}^{n+1}\Lambda_{ai}^\prime k_{\rho^\prime,i}\to-\infty$ when $c\to\infty$, there exists a positive $c$ and a nonempty subset $T_{\rho,r}^\prime$ of $T_{\rho,r}$ such that
\begin{equation}
\begin{split}
    \sum_{i=0}^{n+1}\Lambda_{ai}^\prime k_{\rho^\prime,i}=b_\rho,&\quad a\in T_{\rho,r}^\prime\subset T_{\rho,r}~,\\
    \sum_{i=0}^{n+1}\Lambda_{ai}^\prime k_{\rho^\prime,i}>b_\rho,&\quad a\notin T_{\rho,r}^\prime,~a\in T_{\rho,r}~.
\end{split}
\end{equation}
Then we have
\begin{subequations}
    \begin{align}
        \sum_{i=0}^{n+1}\Lambda_{ai}^\prime k_{\rho^\prime,i}=&b_{\rho^\prime}\equiv b_\rho,\quad a\in S_{\rho^\prime}\equiv S_{\rho,r}\cup T_{\rho,r}^\prime~,\\
        \sum_{i=0}^{n+1}\Lambda_{ai}^\prime k_{\rho^\prime,i}>&b_{\rho^\prime},\quad a\notin S_{\rho^\prime}~.
    \end{align}
\end{subequations}
Thus we get a new region $\rho^\prime$. The scaling of the integral $I(\lambda_0,\lambda_1,\ldots,\lambda_n)$ in the region $\rho^\prime$ is
\begin{equation}
    I(\lambda_0,\lambda_1,\ldots,\lambda_n)\sim x_0^{\nu_\rho+c\nu_{\rho,r}}~.
\end{equation}
Since $c\nu_{\rho,r}$ is a rational number, the region $\rho^\prime$ is degenerate with the region $\rho$. We conclude:

{\it If the asymptotic expansion of a well-defined parametric integral in a region results in an integral with unregulated divergences, this region degenerates with another region.}

\section{Reanalysis of the example}\label{sec:ReanExampl}
We reconsider the example shown in sec.~\ref{sec:Exampl}. The $\mathcal{F}$ polynomial for the integral in eq.~(\ref{eq:Exampl}) is
\begin{equation}
    \mathcal{F}=m^2 x_3 \left(x_1+x_2+x_3\right)-2 x_1 x_2+\left(x_1+x_2+x_3\right) x_4~.
\end{equation}
By using the geometric method, three regions can be found for this integral family, which correspond to those in eq.~(\ref{eq:Reg}). The corresponding scaling vectors $\bm{k}_\rho$ are
\begin{equation}
\begin{split}
k_{\rho_1} =& ~(1,~0,~-2,~-2,~0)~, \\
k_{\rho_2} =& ~(1,~-2,~0,~-2,~0)~, \\
k_{\rho_3} =& ~(1,~0,~0,~0,~0)~,
\end{split}
\end{equation}

We consider the first region. The leading terms of this region (that is, terms in $S_{\rho_1}$) are
\begin{equation}
    \mathcal{F}_a=m^2 x_3 \left(x_2+x_3\right)-2 x_1 x_2+\left(x_2+x_3\right) x_4~.
\end{equation}
For the integral $I_a(-\frac{d}{2},~0,~0,~0)$ in this integral family, there is one region with unregulated divergences. The corresponding region vector and scaling are
\begin{equation}
    \begin{split}
        k_{\rho_1,r_1}=&~(-1,~1,~0,~0)~,\\
        \nu_{\rho_1,r_1}=&~0~.
    \end{split}
\end{equation}
This region leads to the degeneracy of the two collinear regions, because
\begin{equation}
\begin{split}
    k_{\rho_2,i} = &~k_{\rho_1,i} + 2 k_{\rho_1,r_1,i},\quad i>0~,\\
    \nu_{\rho_2} = &~\nu_{\rho_1}+2\nu_{\rho_1,r_1}=\nu_{\rho_1}~.
\end{split}
\end{equation}

\section{Discussions}
In this article, we provide a criterion to identify unregulated divergences for Feynman integrals. It is further proven that if a divergence arises from the asymptotic expansion of a Feynman integral in a region, this region must be degenerate with another region.

The analysis is based on the parametric representation, and only nonthreshold regions are considered. The generalization of this analysis to threshold regions~\cite{Jantzen:2012mw, Semenova:2018cwy,Ananthanarayan:2018tog} and phase-space integrals~\cite{Chen:2020wsh} will be considered in the future. It would also be interesting to provide a parallel analysis in the momentum space~(see e.g.~ref.~\cite{Gardi:2022khw}), and consider the consequences of the degeneracies of regions on effective theories and the factorizations of amplitudes.

\section*{Acknowledgement}
This work is supported by Guangdong Major Project of Basic and Applied Basic Research~(No. 2020B0301030008).

\bibliographystyle{apsrev4-2} 
\bibliography{refs}






\end{document}